\documentclass[3p,times,procedia]{elsarticle}
\usepackage{nupha_ecrc}

\newcommand{\Jpsi}{$J/\psi$ }
\newcommand{\pT}{$p_{_T}$ }

\newcommand{\sNN}{$\sqrt{s_{_\mathrm{NN}}}$ }
\newcommand{\s}{$\sqrt{s}$ }
\newcommand{\pp}{$p$+$p$ }

\newcommand{\auau}{Au+Au }

\volume{00}

\firstpage{1}

\journalname{Nuclear Physics A}

\runauth{}

\jid{nupha}

\jnltitlelogo{Nuclear Physics A}

\usepackage{amssymb}

\usepackage[figuresright]{rotating}

\begin{document}

\begin{frontmatter}



\dochead{XXVIIIth International Conference on Ultrarelativistic Nucleus-Nucleus Collisions\\ (Quark Matter 2019)}

\title{Quarkonium production: An experimental overview}


\author[sklpde,mphy] {Zebo Tang}
\address[sklpde]{State Key Laboratory of Particle Detection and Electronics, University of Science and Technology of China, Hefei, Anhui, China}
\address[mphy]{Department of Modern Physics, University of Science and Technology of China, Hefei, Anhui, China}

\begin{abstract}
Quarkonium has been proposed as a sensitive probe of quark-gluon plasma (QGP) more than thirty years ago. Since then, lots of experimental efforts have been devoted to study its production in heavy-ion collisions to search for QGP and study its properties and significant progresses have been made. In this paper, an overview of recent experimental results on charmonium and bottomonium production in heavy-ion collisions as well as in small systems are presented. Furthermore, the results on exotic particle X(3872) production in Pb+Pb and \pp collisions are also discussed. 
\end{abstract}

\begin{keyword}
Quarkonium, Heavy Flavor, Relativistic Heavy-ion collisions, Quark-Gluon Plasma

\end{keyword}
\end{frontmatter}


\section{Introduction}
\label{}

Quarkonium is a tightly bound state of a heavy quark and its anti-quark. Charmonium (bottomonium) refers to the bound state of charm (bottom) quark and its anti-quark. Table~\ref{tab:quarkonium} shows the mass, binding energy and radius of various quarkonium states~\cite{quarkonium_size_Satz06}. Although the production mechanism in elementary collisions such as \pp collisions is still not fully understood, quarkonium has been proposed as a unique probe of the deconfined hot and dense medium, so-called quark-gluon plasma (QGP), created in ultra-relativistic heavy-ion collisions by T. Matsui and H. Satz in 1986~\cite{colorscreen}. If QGP is formed in heavy-ion collisions, the production yield of quarkonium is expected to be significantly suppressed with respect to the yield in \pp collisions scaled by the number of binary nucleon-nucleon collisions because the potential of the heavy quark and its anti-quark is modified by the deconfined medium and quarkonium states are subject to be dissociated (or melted) when the temperature of the medium is high enough. The temperature required to dissociate a quarkonium state (dissociation temperature, $T_d$) depends on the binding energy or radius of the quarkonium state. More loosely bound state (lower binding energy or larger radius) has lower $T_d$. In both charmonium and bottomonium sectors, $T_d$ decreases with increasing quarkonium mass and the excited states have lower $T_d$ than the ground state. When put the charmonium and bottomonium states together as shown in Tab.~\ref{tab:quarkonium} for comparison, one finds that $T_d^{\Upsilon(1S)} > T_d^{\chi_b} \sim T_d^{J/\psi} \sim T_d^{\Upsilon(2S)} > T_d^{\chi_b'} \sim T_d^{\chi_c} \sim T_d^{\Upsilon(3S)} > T_d^{\psi(2S)}$. The systematical measurements of quarkonium suppression can also help to constrain the temperature profile and the dynamic evolution of the fireball produced in ultra-relativistic heavy-ion collisions.

However, other effects need to be considered. First of all, in the deconfined medium, the (un)correlated heavy quark and anti-quark could (re)combine into a quarkonium state when they get close enough in phase space.  The probability is proportional to the square of the total number of heavy quark and anti-quark pairs produced in one collision. The production yield of quarkonium originated from (re)combination is expected to be enhanced in central heavy-ion collisions with respect to peripheral heavy-ion collisions and \pp collisions. Although the (re)combination effect is competing with the QGP melting effect, both of them require deconfinement and can be used to search for QGP and study its properties. In addition to these two hot nuclear matter effects, quarkonium production in heavy-ion collisions is also affected by cold nuclear matter (CNM) effects, including modification of parton distribution function in nuclei (nPDF), breakup by hadrons, the scattering and/or energy loss of the parton evolved in quarkonium production, etc. The CNM effects can be experimentally studied via the collisions of $p$ or light nucleus and heavy nucleus. There are other effects need to be considered when interpreting the experimental results. One important effect is the feed-down contribution of quarkonium production. For example, the inclusive \Jpsi consists of prompt \Jpsi and non-prompt $J/\psi$. The latter is referring to the contribution from the decay of $B$-hadrons. While the former includes direct \Jpsi and the feed-down from excited charmonium states. The composition of different prompt quarkonium states can be found at~\cite{quarkonium_feeddown}. The measured suppression for certain (inclusive or prompt) quarkonium state is not purely due to the suppression of the directly produced quarkonium state, but has also contribution from the suppression of various feed-down sources. Another important effect is the contribution from jet fragmentation at intermediate and high \pT range. The quarkonium from jet fragmentation could form outside of the fireball thus is not affected by the QGP melting or (re)combination, but affected by ``jet quenching''.

In the following sections, the selected latest experimental results obtained at RHIC and LHC will be presented and physics implications will be discussed.


\begin{table}[tb]\centering
\caption{The mass, binding energy and radius of charmonium and bottomonium states.}
\begin{tabular}{c|ccc|ccccc}
\hline
State & \multicolumn{3}{c|}{Charmonium} & \multicolumn{5}{c}{Bottomonium} \\
{} & \Jpsi & $\chi_c$ & $\psi(2S)$ 
& $\Upsilon(1S)$ & $\chi_b$ & $\Upsilon(2S)$ & $\chi_b'$  & $\Upsilon(3S)$  \\
\hline
Mass (GeV/$c^2$) & 3.10 & 3.53 & 3.68 & 9.46 & 9.99 & 10.02 & 10.26 & 10.36 \\
$\Delta E$ (GeV/$c^2$) & 0.64 & 0.20 & 0.05 & 1.10 & 0.67 & 0.54 & 0.31 & 0.20 \\
Radius (fm) & 0.25 & 0.36 & 0.45 & 0.14 & 0.22 & 0.28 & 0.34 & 0.39 \\
\hline
\end{tabular}
\label{tab:quarkonium}
\end{table}

%
%

\section{Quarkonium production in $p$+A collisions and small systems}

\begin{figure}[!htb]\centering
\includegraphics [width=0.95\hsize] 
{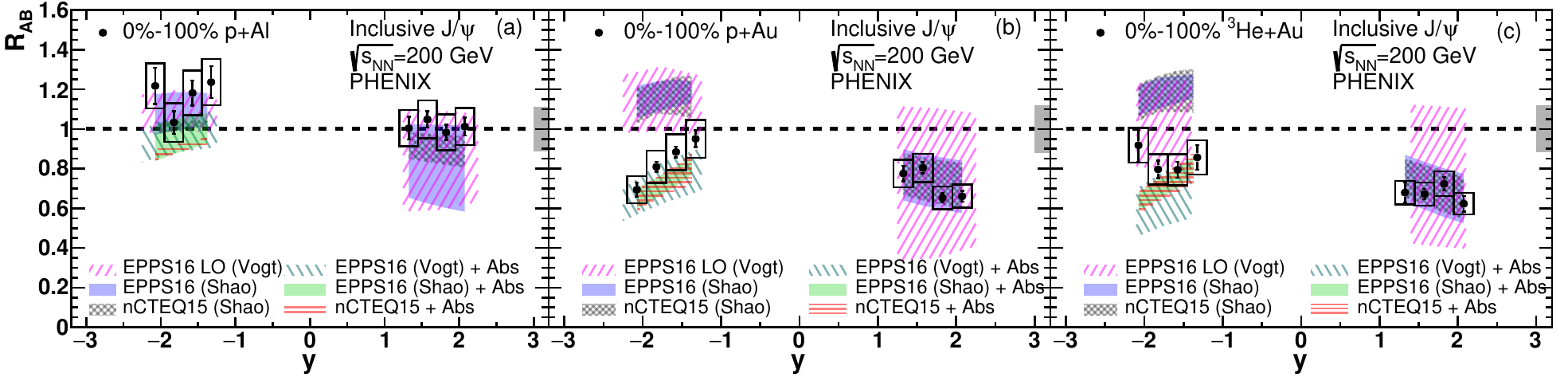}
\caption{\Jpsi nuclear modification factor as a function of rapidity in $p$+Al, $p$+Au and $^3$He+Au collisions at \sNN = 200 GeV~\cite{PHENIX_Jpsi_pAlpAuHeAu}.}
\label{fig:Jpsi_pA}
\end{figure}

The PHENIX Collaboration has measured inclusive \Jpsi nuclear modification factor as a function of rapidity at forward ($p$/$^3\textrm{He}$-going direction) and backward (Al/Au-going direction) rapidity in $p$+Al, $p$+Au and $^3\textrm{He}$+Au collisions at \sNN = 200 GeV as shown in Fig.~\ref{fig:Jpsi_pA}~\cite{PHENIX_Jpsi_pAlpAuHeAu}. The theoretical calculations with nPDF only and with nPDF plus nuclear absorption are also shown for comparison. At forward rapidity ($p$/$^3$He-going direction), significant suppression for inclusive \Jpsi is observed for Au target but consistent with no suppression for Al target. The theoretical calculations incorporating nPDF (EPPS16, nCTEQ15) with and without nuclear absorption describe the data reasonably well. At backward rapidity (Al/Au-going direction), the nuclear modification factor is systematically larger than unity for Al target but exhibits obvious suppression for Au target. The theoretical calculations with nPDF only predict enhancement at backward rapidity, which is higher than the data for Au target. With the nuclear absorption from global fit to world data added, the theoretical calculations can describe the rapidity dependence of the inclusive \Jpsi suppression in $p$+Al, $p$+Au and $^3\textrm{He}$+Au collisions at RHIC reasonably well. It is also found that the suppression in 0-20\% $^3\textrm{He}$+Au collisions is very similar as in $p$+Au, suggesting that there is little final state effect on \Jpsi production in $p$+Au and $^3\textrm{He}$+Au collisions.

\begin{figure}[!htb] \centering{
\includegraphics 
  [width=0.38\hsize]
  {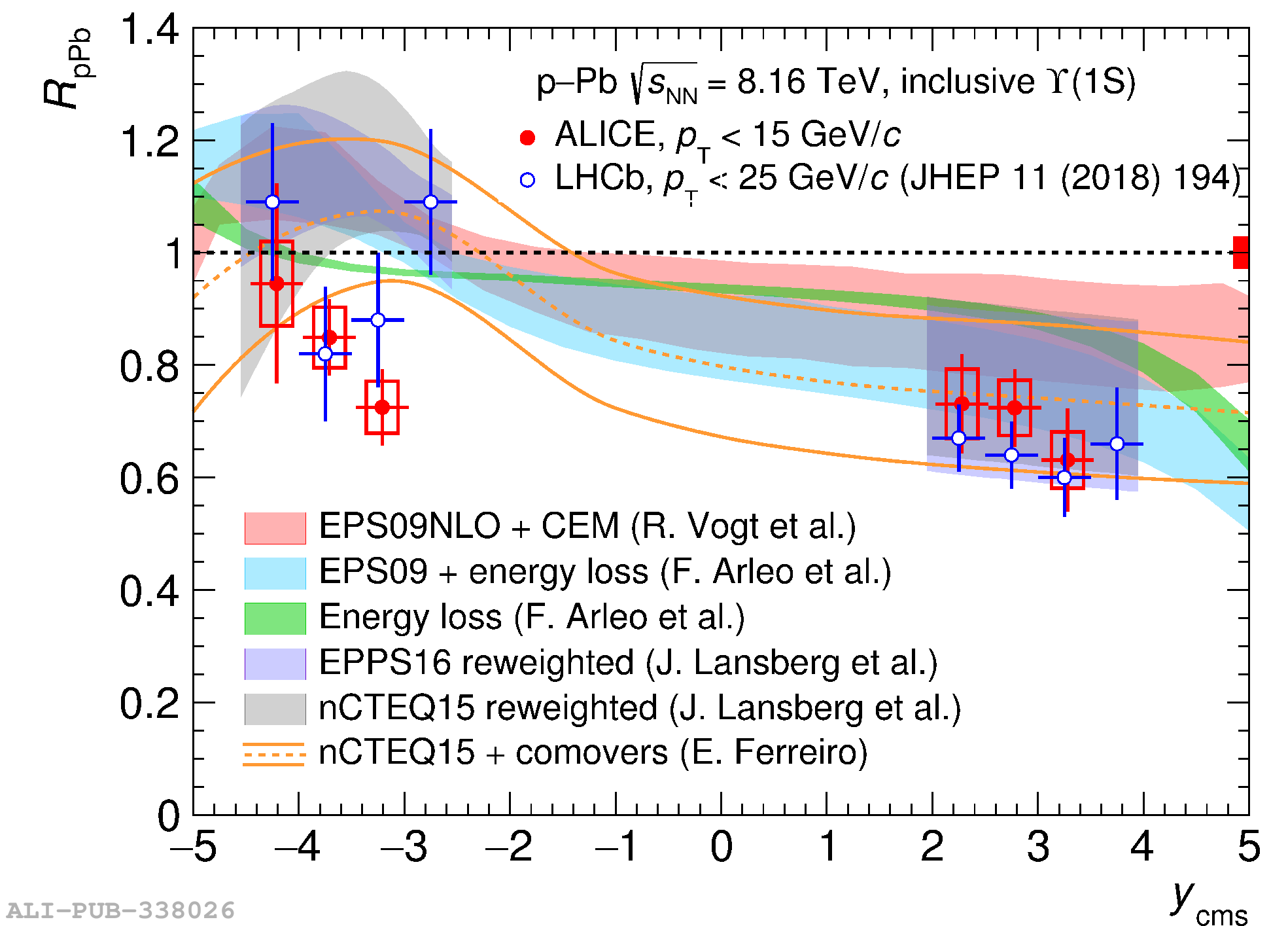}
  \includegraphics
  [width=0.3\hsize]
  {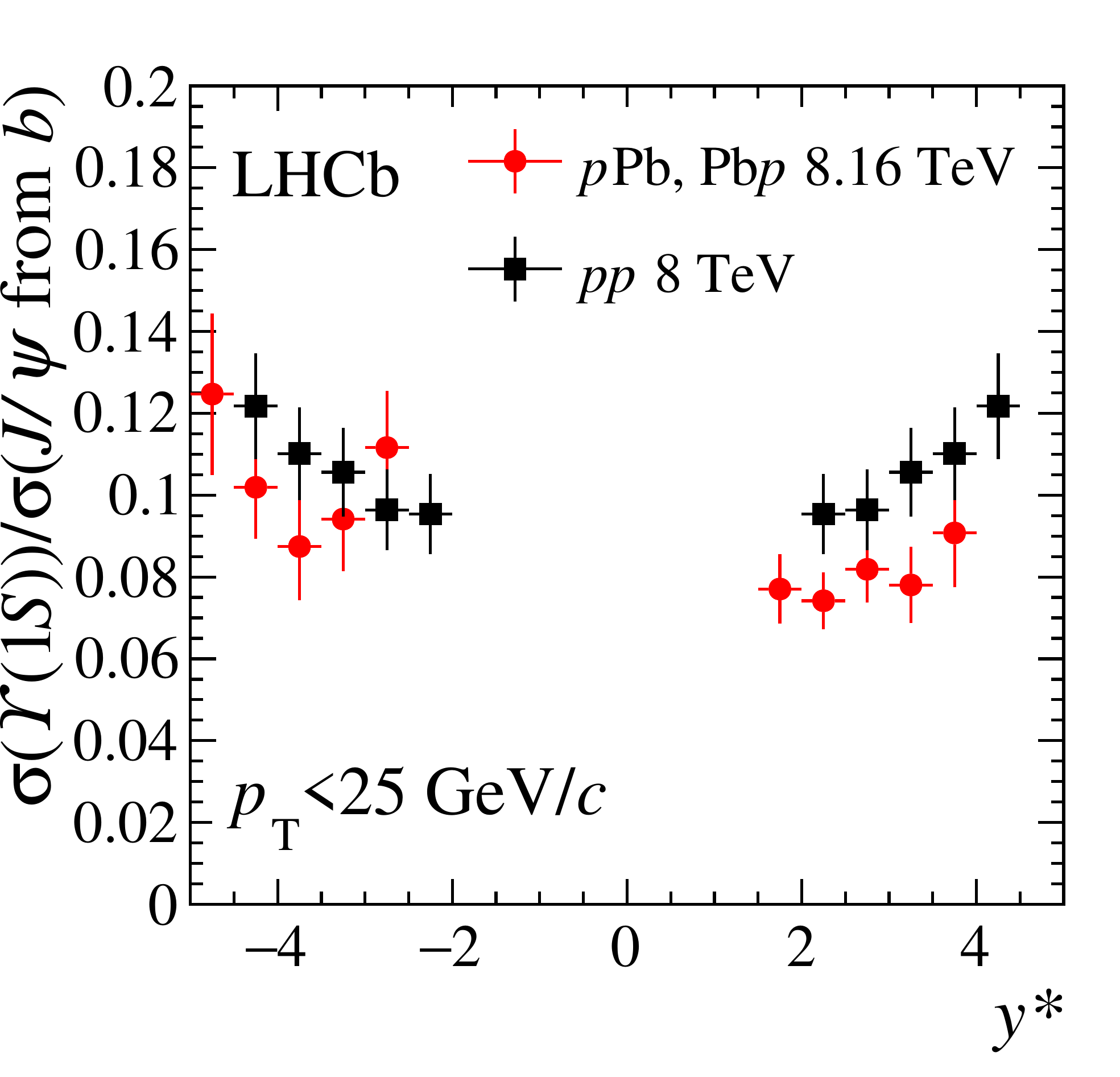}
  \includegraphics
    [width=0.3\hsize]
  {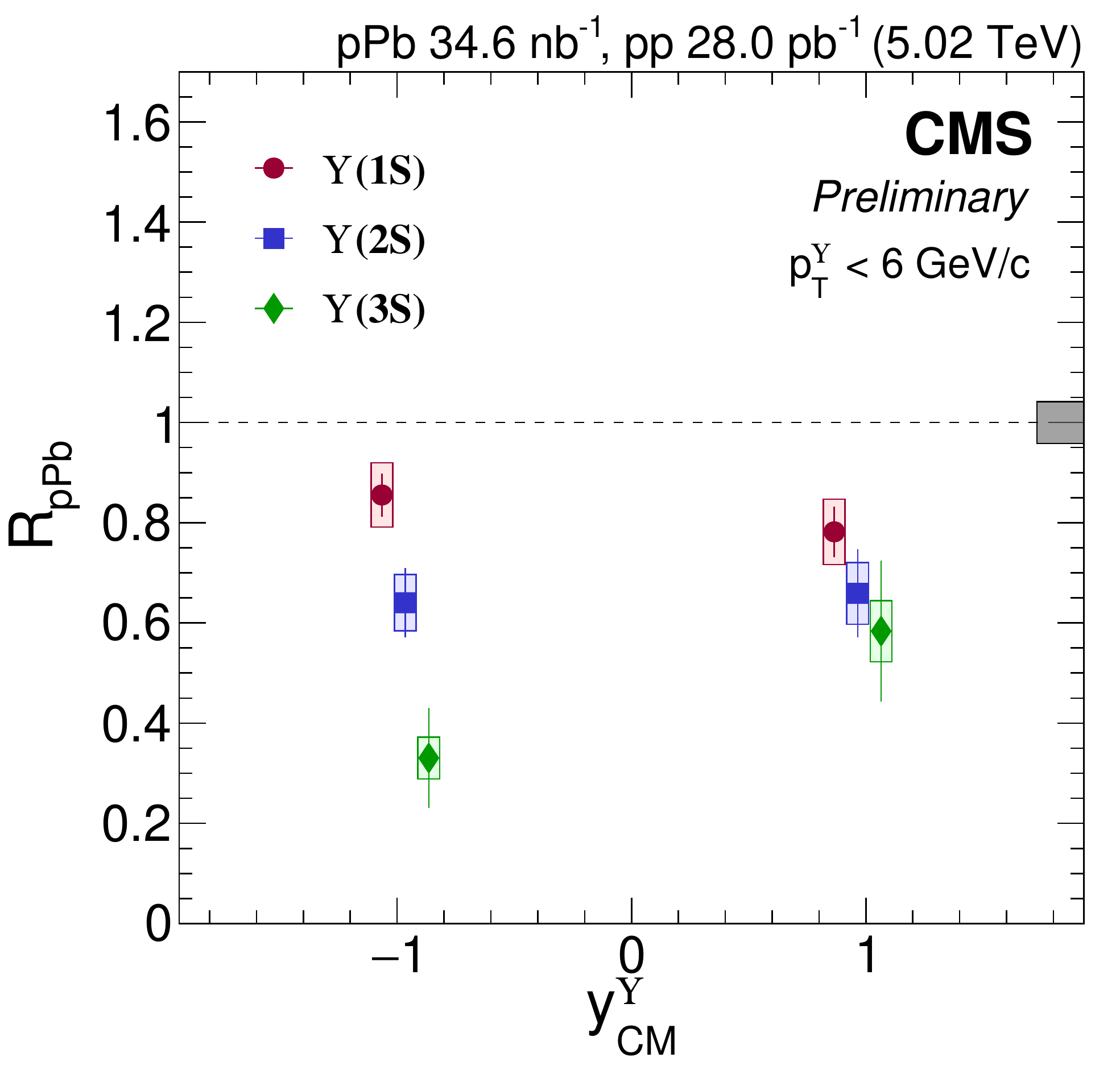}}
\caption{Left: Inclusive $\Upsilon(1S)$ nuclear modification factor as a function of rapidity in $p$+Pb collisions at \sNN = 8.16 TeV~\cite{Y1S_pPb_ALICE}. Middle: The cross section ratio of $\Upsilon(1S)$ and \Jpsi from $b$ decays in \pp and $p$+Pb collisions~\cite{Y1S_pPb_LHCb}. Right: The nuclear modification factor of $\Upsilon(1S, 2S, 3S)$ as a function of rapdity in $p$+Pb collisions at \sNN = 5.02 TeV~\cite{Y1S2S3S_pPb_CMS}.}
\label{fig:Y1S_pPb}
\end{figure}

The ALICE Collaboration has measured inclusive $\Upsilon(1S)$ nuclear modification factor in $p$+Pb collisions at \sNN = 8.16 TeV, as a function of rapidity, in the forward and backward rapidity regions as shown in the left panel of Fig.~\ref{fig:Y1S_pPb}~\cite{Y1S_pPb_ALICE}. The LHCb results are also shown for comparison~\cite{Y1S_pPb_LHCb}. At forward rapidity, strong suppression is observed and can be described by theoretical calculations with various nPDF only, energy loss only, nPDF plus energy loss, nPDF plus breakup by comovers, etc. At backward rapidity, the data is systematically lower than unity and the theoretical calculations. LHCb also measured the production cross section ratio of inclusive $\Upsilon(1S)$ and \Jpsi from $b$ decays as a function of rapidity in \pp collisions at \s = 8 TeV and $p$+Pb collisions at \sNN = 8.16 TeV as shown in the middle panel of Fig.~\ref{fig:Y1S_pPb}~\cite{Y1S_pPb_LHCb}. At backward rapidity, the ratios in \pp and $p$+Pb are consistent with each other. While at forward rapidity ($p$-going direction), a small suppression of the ratio is visible in $p$+Pb collisions with respect to \pp collisions. This indicates different suppression mechanism for bottomonium and open beauty and the physics origin is still under investigation. The right panel of Fig.~\ref{fig:Y1S_pPb} shows the nuclear modification factor of the 1S, 2S and 3S state of $\Upsilon$ in $p$+Pb collisions at \sNN = 5.02 TeV measured by the CMS Collaboration~\cite{Y1S2S3S_pPb_CMS}. The sequential suppression is observed and more pronounced at backward rapidity (Pb-going direction). The nPDF effect on the 1S, 2S and 3S state should be similar and results in $>\sim 1$ nuclear modification factor at backward rapidity. The sequential suppression at backward indicates there is significant final state effects such as breakup by co-movers or possible hot nuclear medium effects.

\section{Quarkonium production in heavy-ion collisions}
\subsection{Charmonium}

\begin{figure}[tb] \centering{
\includegraphics
  [width=0.32\hsize]
  {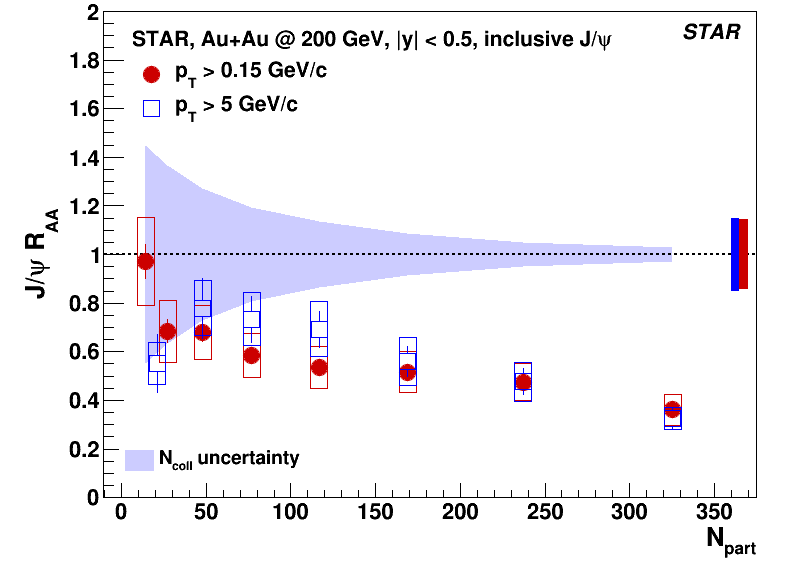}
\includegraphics
  [width=0.32\hsize]
  {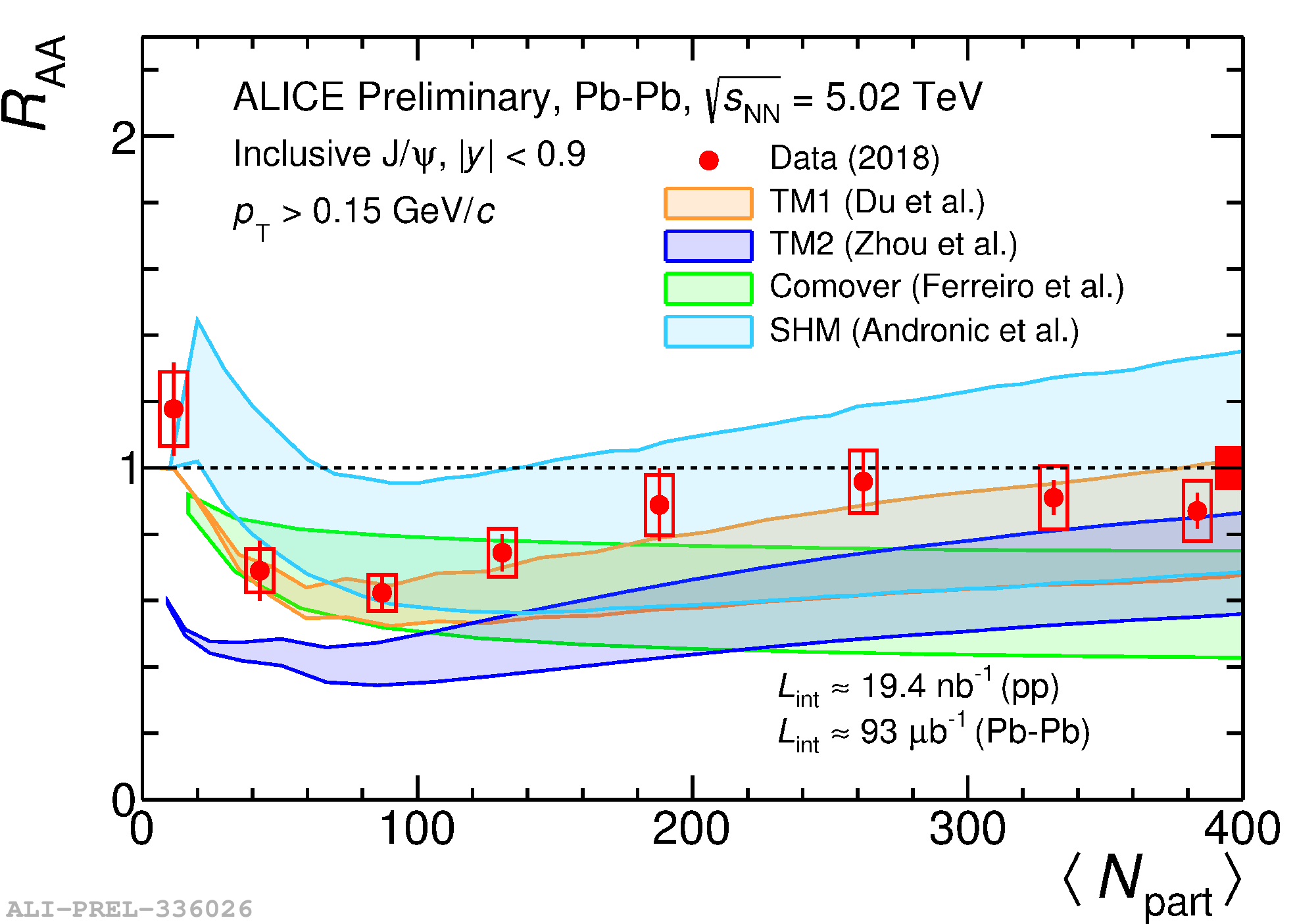}
  \includegraphics
  [width=0.32\hsize]
  {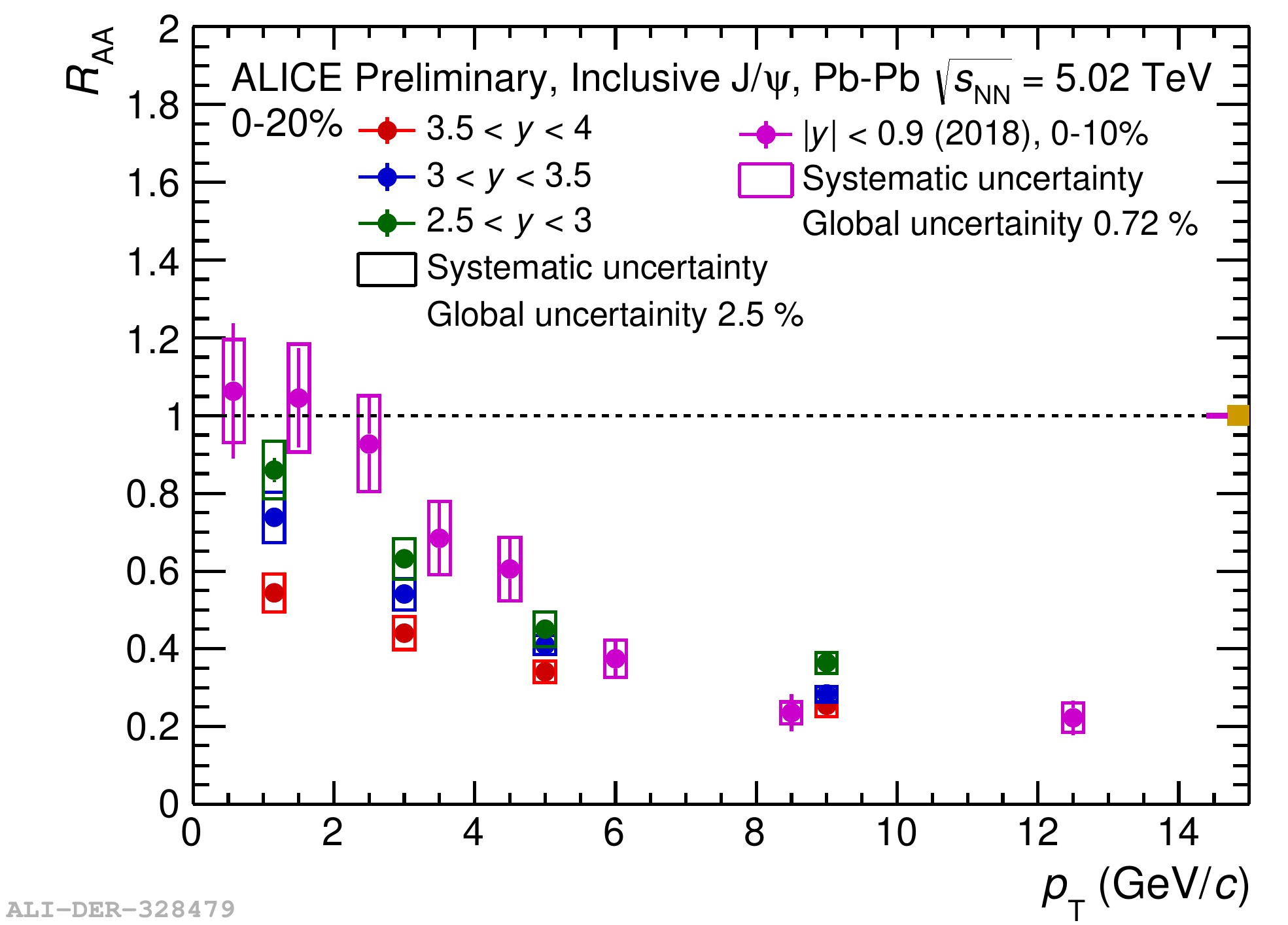}}
\caption{Nuclear modification factor of inclusive \Jpsi in Au+Au collisions at 0.2 TeV~\cite{Jpsi_AuAu_STAR_PLB2019} and Pb+Pb collisions at 5.02 TeV~\cite{XiaozhiBai_QM2019Proceedings}. }
\label{fig:Jpsi_RAA}
\end{figure}

The STAR Collaboration recently published the centrality and \pT dependence of inclusive \Jpsi production in a wide \pT range in \auau collisions at \sNN = 200 GeV~\cite{Jpsi_AuAu_STAR_PLB2019}. The left panel of Fig.~\ref{fig:Jpsi_RAA} shows the inclusive \Jpsi nuclear modification factor as a function of $N_{\mathrm{part}}$ in \auau collision at \sNN = 200 GeV in low-\pT ($p_T>0.15~\textrm{GeV}/c$) and high-\pT ($p_T>5~\textrm{GeV}/c$) regions. The suppression for both low-\pT and high-\pT \Jpsi increases towards central collisions and exhibits significant suppression in central collisions (a factor of $\sim 3$). For low-\pT $J/\psi$, the suppression is due to the interplay of CNM effects, QGP melting and (re)combination effects. While for high-\pT $J/\psi$, the (re)combination effect is negligible. The nuclear modification factor measured in $d$+Au and $p$+Au collisions is found to be consistent with unity at \pT above 5 GeV/$c$, suggesting that CNM effects are also not important at high-$p_T$~\cite{Jpsi_dAu_PHENIX2013, Jpsi_pAu_STARQM2017}. The observed significant suppression of high-\pT \Jpsi provides strong evidence for the QGP melting of charmonium. 

The ALICE Collaboration recently released the results of inclusive \Jpsi nuclear modification factor using the data of Pb+Pb collisions at \sNN = 5.02 TeV taken in 2018~\cite{XiaozhiBai_QM2019Proceedings}. The new results are found to be consistent with the previous results using data taken in 2015, but the precision is significantly improved. The middle panel of Fig.~\ref{fig:Jpsi_RAA} shows the centrality dependence of the inclusive \Jpsi nuclear modification factor at mid-rapidity. Unlike the decreasing trend observed at RHIC energy, the nuclear modification factor at LHC decreases from peripheral collisions to semi-peripheral collisions then increases and saturates at around unity in semi-central and central collisions. The nuclear modification factor in central collisions at LHC is much larger than at RHIC energy for low-\pT $J/\psi$. The \pT dependence of the nuclear modification factor is also very different at RHIC and LHC. At RHIC, the \pT dependence is rather flat, but there is clear decreasing trend with increasing \pT at LHC as shown in the right panel of Fig.~\ref{fig:Jpsi_RAA}. These results provide strong evidence of significant contribution of (re)combination for low-\pT \Jpsi at LHC energy. 
The right panel of Fig.~\ref{fig:Jpsi_RAA} compares the \pT dependence of the inclusive \Jpsi nuclear modification factor in various rapidity windows. In all of the presented rapidity windows, the nuclear modification factor shows decreasing trend. At low $p_T$, the nuclear modification factor decreases with increasing rapidity. This is also 
consistent with the picture that (re)combination plays dominant role for low-\pT \Jpsi production at LHC with the fact that the charm quark production cross section decreases with increasing rapidity. At high $p_T$ region, the rapidity dependence is found to be more complicated. The Statistical Hadronization Model and Transport Models can describe the centrality dependence of the inclusive \Jpsi nuclear modification factor within uncertainties. But the uncertainties of the theoretical calculations are larger than data, suggesting that more precise measurements on charm cross section and CNM effects are needed to provide better constraint. 

\begin{figure}[!htb] \centering{
\includegraphics
  [width=0.35\hsize]
  {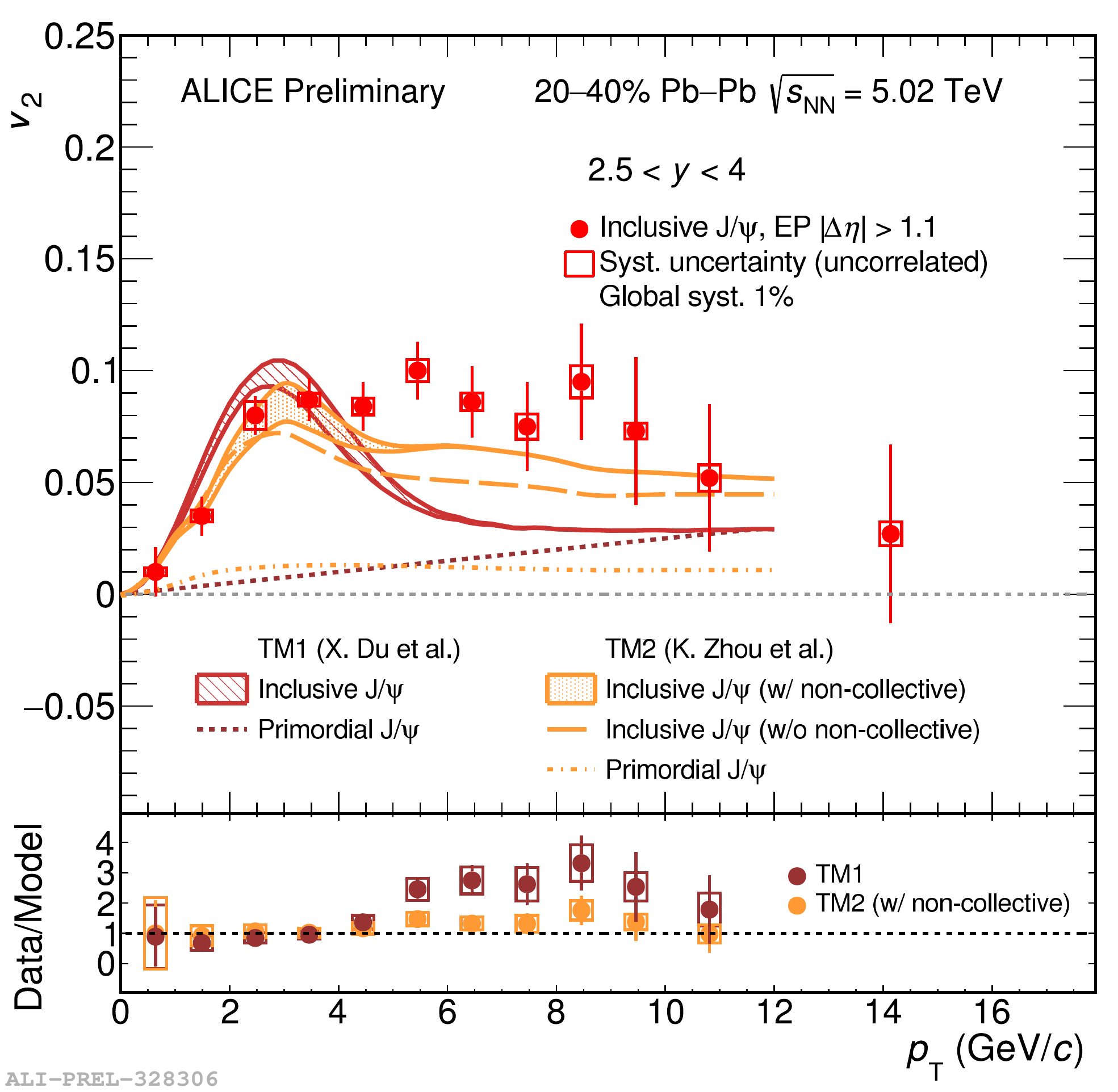}
\includegraphics
  [width=0.38\hsize]
  {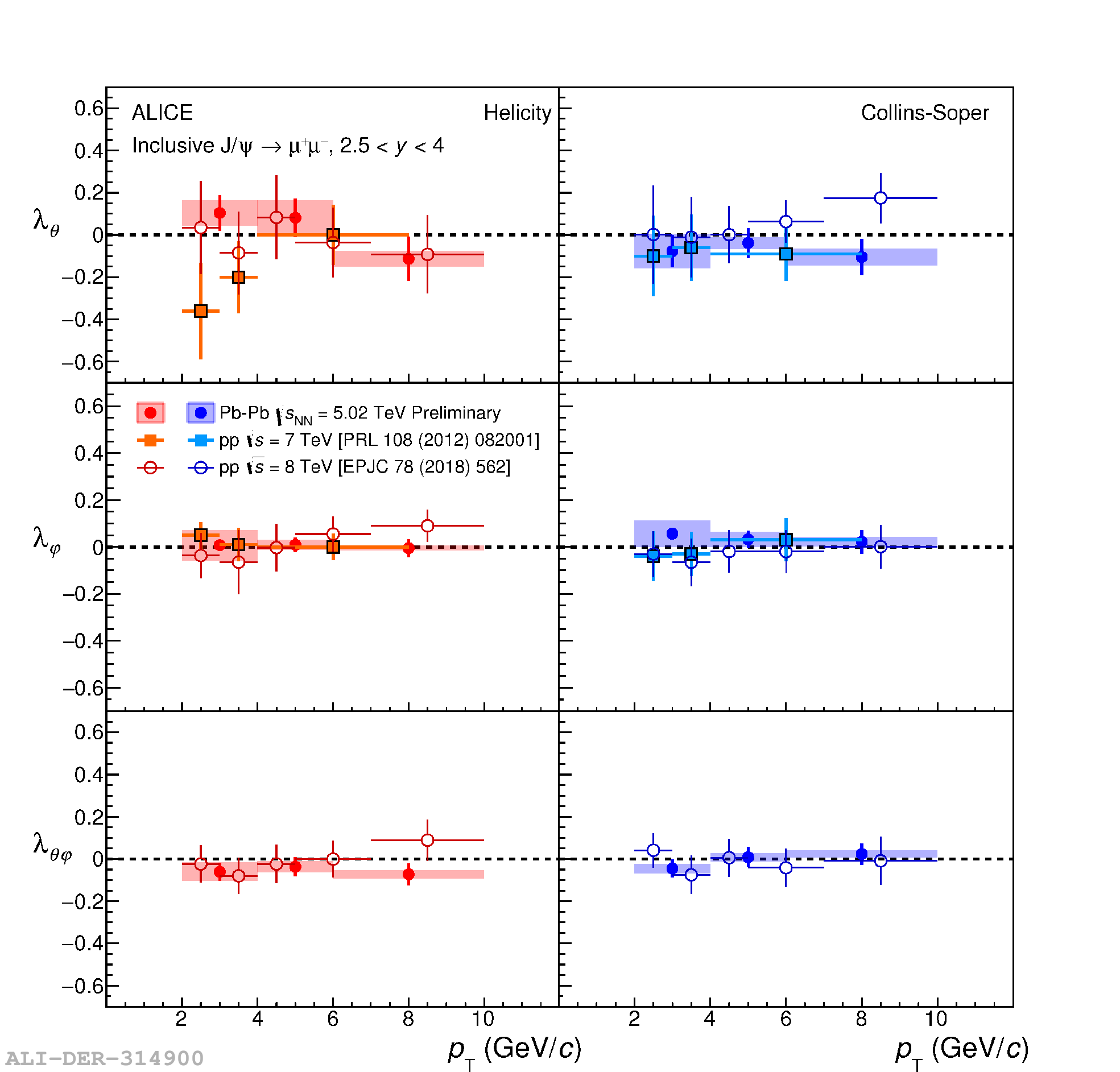}}
\caption{Inclusive \Jpsi elliptic flow $v_2$ (left) and polarization parameters (right) as a function of \pT in Pb+Pb collision at \sNN = 5.02 TeV measured by the ALICE Collaboration~\cite{XiaozhiBai_QM2019Proceedings}. Theoretical calculations on $v_2$ in Pb+Pb collisions and measurements of polarization in \pp collisions are also shown for comparison.}
\label{fig:Jpsi_v2_Pol}
\end{figure}

The (re)combination effect can also be investigated via the measurements of quarkonium elliptic flow $v_2$. The experiments at both RHIC and LHC found $v_2$ of open charm mesons follow the number-of-constitute-quark (NCQ) scaling as the light flavors, suggesting that charm quarks are thermalized and gain flow via the interaction with the hot, dense medium. Quarkonium produced via the heavy quark (re)combination should inherit the flow of the heavy quarks.  The ALICE Collaboration has measured the inclusive \Jpsi $v_2$ as a function of \pT at forward rapidity as shown in the left panel of Fig.~\ref{fig:Jpsi_v2_Pol} and mid-rapdity (not shown) in Pb+Pb collisions at \sNN = 5.02 TeV~\cite{XiaozhiBai_QM2019Proceedings}. Significant positive $v_2$ is observed. The data at low-\pT can be described by transport model calculations, indicating that charm quarks are thermalized and (re)combination is the dominant contribution for low-\pT J/$\psi$. At \pT above 4 GeV/$c$, the data is systematically higher than the model calculations even with the non-prompt \Jpsi contribution being taken into account. Other effect such as contribution from jet fragmentation probably need to be considered. Based on the measurements of \Jpsi production in a jet in \pp collisions at 8 TeV, the CMS Collaboration concluded that ``jet fragmentation is the dominate source of prompt \Jpsi mesons with $E_{J/\psi} > 15$ GeV and $|y_{J/\psi}|<1$''~\cite{Jpsi_jet_fragmentation_pp8TeV_CMS}. It is of particular interest to push the measurement to lower $E$ or $p_T$ to study the fraction of jet fragmentation at lower $p_T$. 

The ALICE Collaboration also reported the first measurement of \Jpsi polarization in Pb+Pb collisions~\cite{XiaozhiBai_QM2019Proceedings}. \Jpsi polarization in heavy-ion collisions provides a novel probe of quarkonium production mechanism in QGP. \Jpsi polarization in heavy-ion collisions could be modified with respect to \pp collisions by various effects: 1) Quarkonium produced via (re)combination should be different from that of primordial $J/\psi$; 2) Since different quarkonium states are expected to have different polarization, sequential evolution of quarkonium polarization is expected due to sequential suppression; 3) Possible modification by strong electromagnetic field; 4) The screening of non-perturbative effects in quarkonium production in QGP may result in increase of quarkonium polarization~\cite{Jpsi_Pol_Dima_PRC03}; and so on. With current uncertainties, the ALICE results in Pb+Pb collisions are consistent with no polarization and \pp results. The precision is expected to be significantly improved with LHC Run-3 data.

\subsection{Bottomonium}

Due to much lower production cross section of $b\bar{b}$ than that of $c\bar{c}$, bottomonium is expected to be less affected by (re)combination effect thus provide a better probe of QGP melting. Recently, the measurements of $\Upsilon(1S, 2S, 3S)$ suppression in Pb+Pb collisions at \sNN = 5.02 TeV are available (ATLAS~\cite{Upsilon_PbPb_ATLASQM2019} and CMS~\cite{Upsilon_PbPb5TeV_CMS}) at mid-rapidity and ALICE~\cite{XiaozhiBai_QM2019Proceedings} at forward rapidity). Significant suppression of $\Upsilon(1S)$ is observed and the rapidity dependence seems to be more flat than that at 2.76 TeV. The sequential suppression in the bottomonium sector observed at 2.76 TeV is also confirmed with the data at 5.02 TeV with improved precision. 

\begin{figure}[htb] \centering{
\includegraphics
  [width=0.3\hsize, height=0.27\hsize]
  {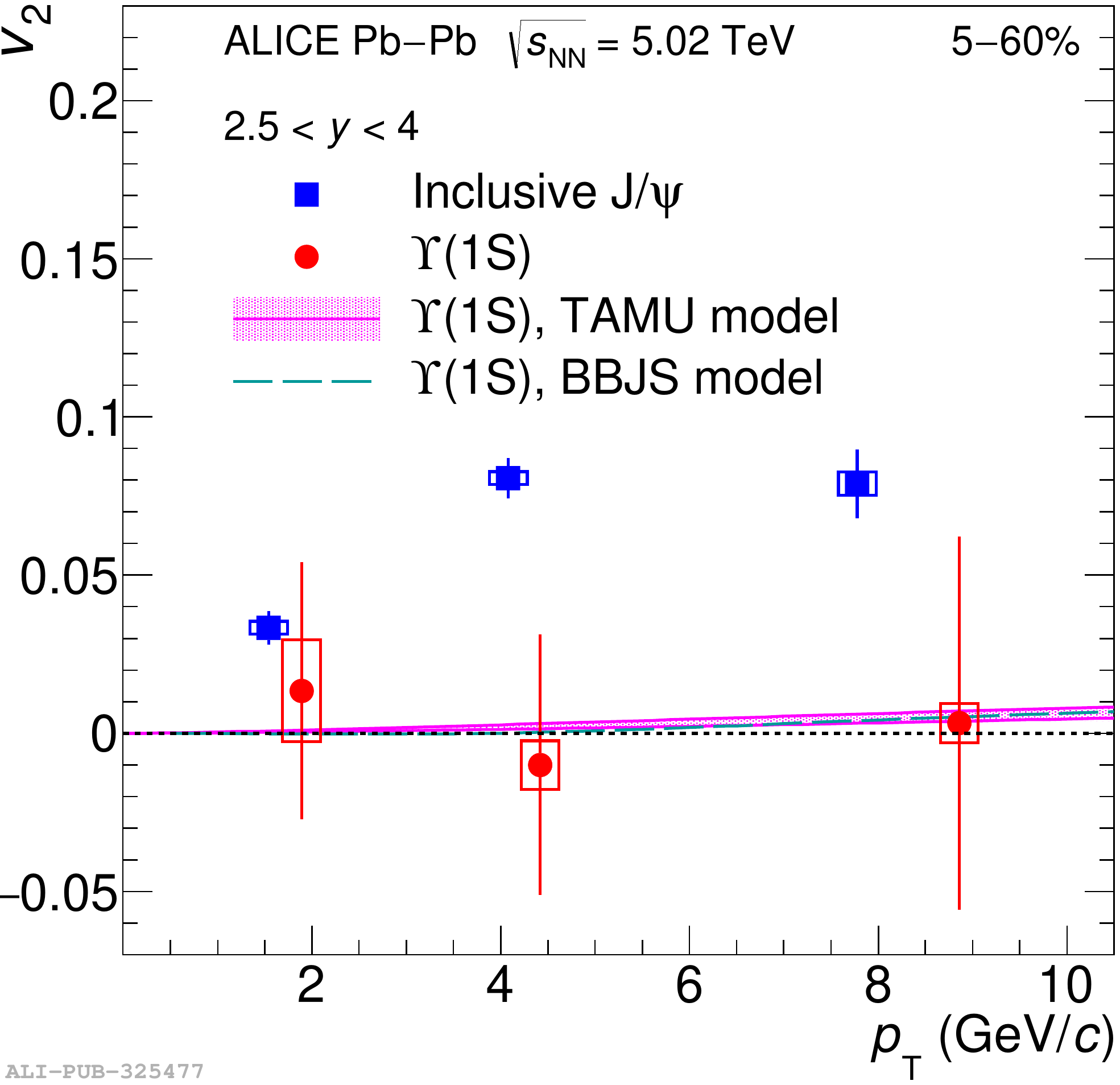}
  \includegraphics
  [width=0.34\hsize, height=0.29\hsize]
  {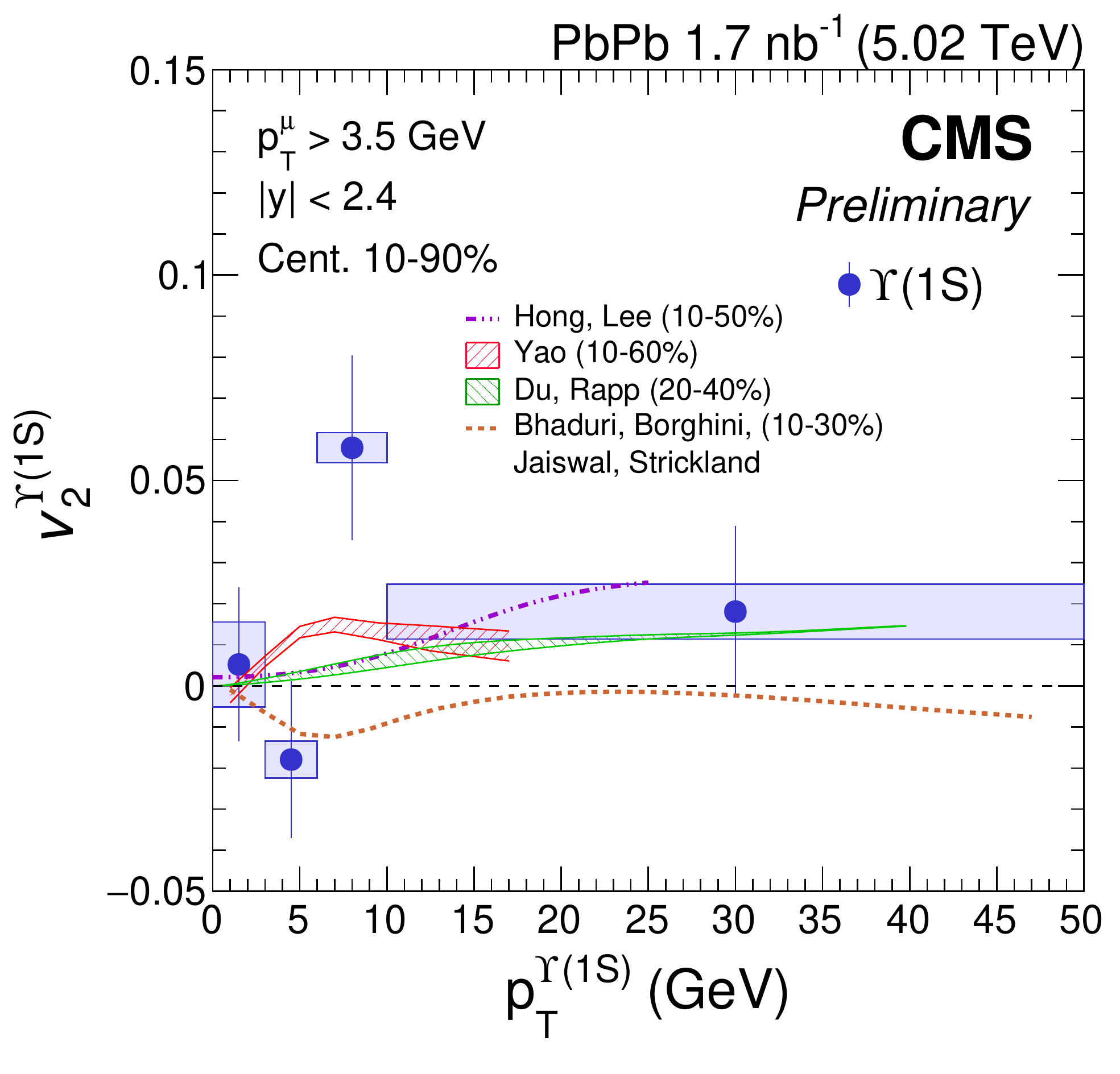}}
\caption{$\Upsilon(1S)$ $v_2$ at forward rapidity (left) and mid-rapidity (right) in Pb+Pb collisions at \sNN = 5.02 TeV measured by ALICE~\cite{ALICE_Upsilon_v2} and CMS~\cite{CMS_Upsilon_v2}, respectively.}
\label{fig:Y1S_v2}
\end{figure}

Lots of attention was drawn by the first measurement of $\Upsilon(1S)$ $v_2$ in heavy-ion collisions done by the ALICE~\cite{ALICE_Upsilon_v2}  and CMS~\cite{CMS_Upsilon_v2} Collaborations. Figure~\ref{fig:Y1S_v2} shows the data and the comparison to inclusive \Jpsi and theoretical calculations. The $v_2$ of $\Upsilon(1S)$ is significantly lower than that of inclusive \Jpsi and consistent with zero and theoretical calculations within large uncertainties on the measurements. For the physics interpretation of the data, there are several questions need to be understood: 1) Does bottom quark flow? The non-zero $v_2$ of $b$-hadrons has been observed, but it does not mean bottom quark flows; 2) What is the fraction of (re)combination for $\Upsilon(1S)$. 
On the experimental side, it is essential to improve the precision, and push the measurement to higher $p_T$ to enhance the sensitivity to bottom quark flow and the level of (re)combination contribution~\cite{Upsilon_v2_thermal}. 

\subsection{Comparison of \Jpsi and $\Upsilon$}

\begin{figure}[!htb] \centering{
  \includegraphics
  [width=0.33\hsize, height=0.28\hsize]
  {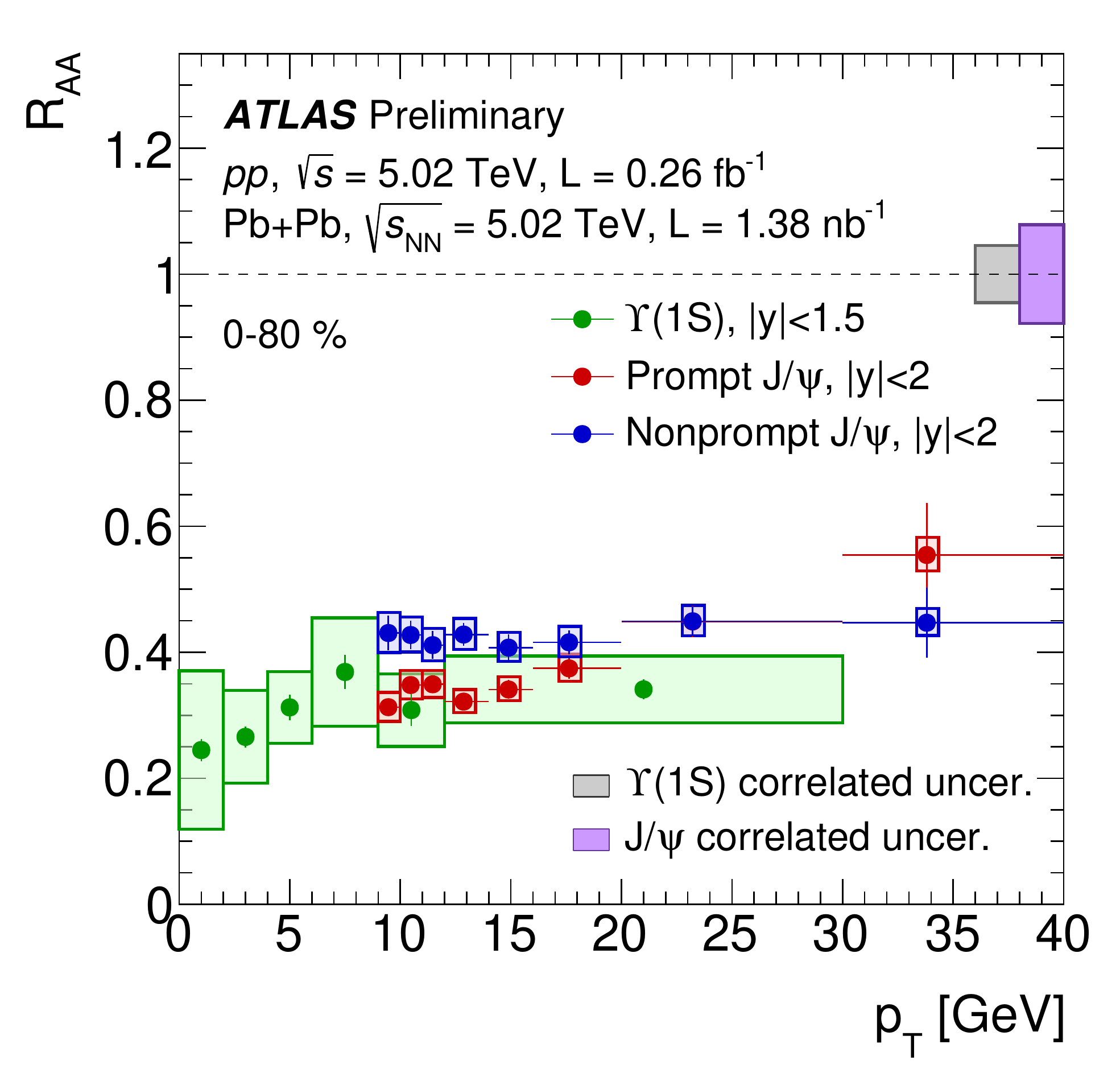}
   \includegraphics
  [width=0.33\hsize, height=0.28\hsize]
  {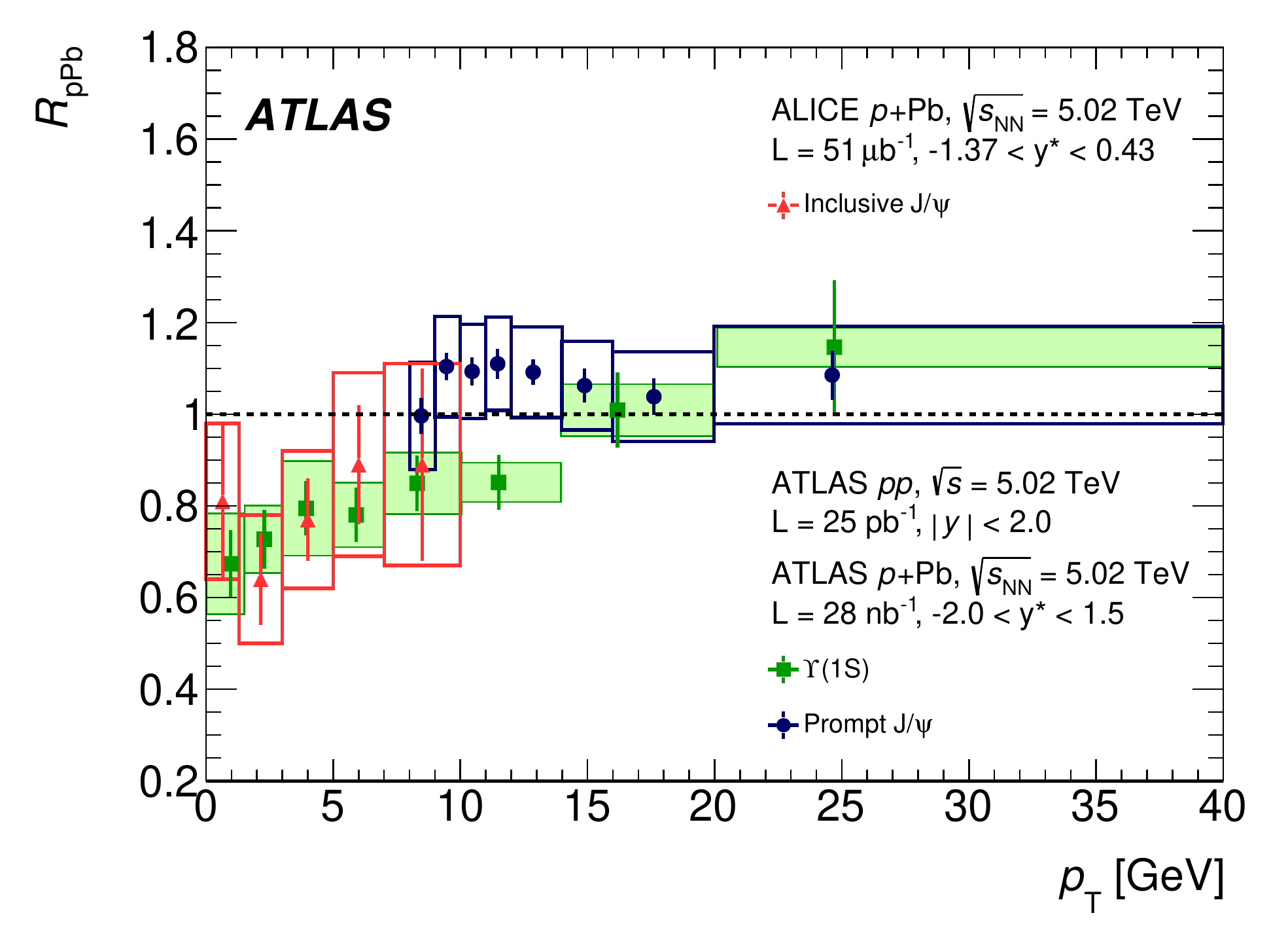}}
\caption{(Color online) \Jpsi and $\Upsilon(1S)$ nuclear modification factor as a function of \pT in Pb+Pb (left) and $p$+Pb (right) collisions at \sNN = 5.02 TeV~\cite{Upsilon_PbPb_ATLASQM2019, quarkonium_pPb_ATLAS2018}. }
\label{fig:Jpsi_Y1S}
\end{figure}

With the high-quality quarkonium data in heavy-ion collisions obtained at RHIC and LHC, it is possible to do comparison between the charmonium sector and bottomonium sector to better understand the physics behind the measurements. In the charmonium sector, I focus on the high \pT region to avoid the contribution from (re)combination. The left panel of Fig.~\ref{fig:Jpsi_Y1S} shows the nuclear modification factor for $\Upsilon(1S)$ and prompt \Jpsi as a function of \pT in Pb+Pb collisions at \sNN = 5.02 TeV~\cite{Upsilon_PbPb_ATLASQM2019}. The suppression of high-\pT \Jpsi due to QGP melting is expected to be stronger than $\Upsilon(1S)$ because the radius of $\Upsilon(1S)$ is much smaller than that of $J/\psi$. However, the nuclear modification factor of $\Upsilon(1S)$ is measured to be surprisingly similar as prompt \Jpsi and lower than non-prompt $J/\psi$ at high $p_T$. Removing possible (re)combination contribution may result in even stronger suppression for $\Upsilon(1S)$. The right panel of Fig.~\ref{fig:Jpsi_Y1S} shows the nuclear modification factor for $\Upsilon(1S)$ and prompt \Jpsi in $p$+Pb collisions at \sNN = 5.02 TeV~\cite{quarkonium_pPb_ATLAS2018}. At $p_T > 14$ GeV/$c$, the suppression of $\Upsilon(1S)$ and prompt \Jpsi are consistent within uncertainties, suggesting that the CNM effects are similar for $\Upsilon(1S)$ and high-\pT prompt $J/\psi$, thus cannot explain the similarity of the suppression of $\Upsilon$ and high-\pT prompt \Jpsi in heavy-ion collisions. How about the feed-down contribution? According to Ref.~\cite{quarkonium_feeddown}, the composition of high-\pT prompt \Jpsi is about 65\% direct production and 35\% contribution from decay of excited charmonium states. While for high-\pT $\Upsilon(1S)$, the composition is about 45\% direct production and 55\% contribution from decay of $\Upsilon(2S)$ and $\chi_b$. Because $\Upsilon(2S)$, $\chi_b$ and \Jpsi has similar radius as shown in Tab.~\ref{tab:quarkonium}, the comparison of prompt high-\pT \Jpsi and $\Upsilon(1S)$ is more like the comparison of direct $\Upsilon(1S)$ and excited charmonium states (such as $\psi(2S)$ and $\chi_c$). The similarity of high-\pT \Jpsi and $\Upsilon(1S)$ suppression may indicate that there are other effects such as jet fragmentation, formation time effect playing role in the quarkonium production at high $p_T$. 
\begin{figure}[!htb]\centering{
\includegraphics
  [width=0.32\hsize]
  {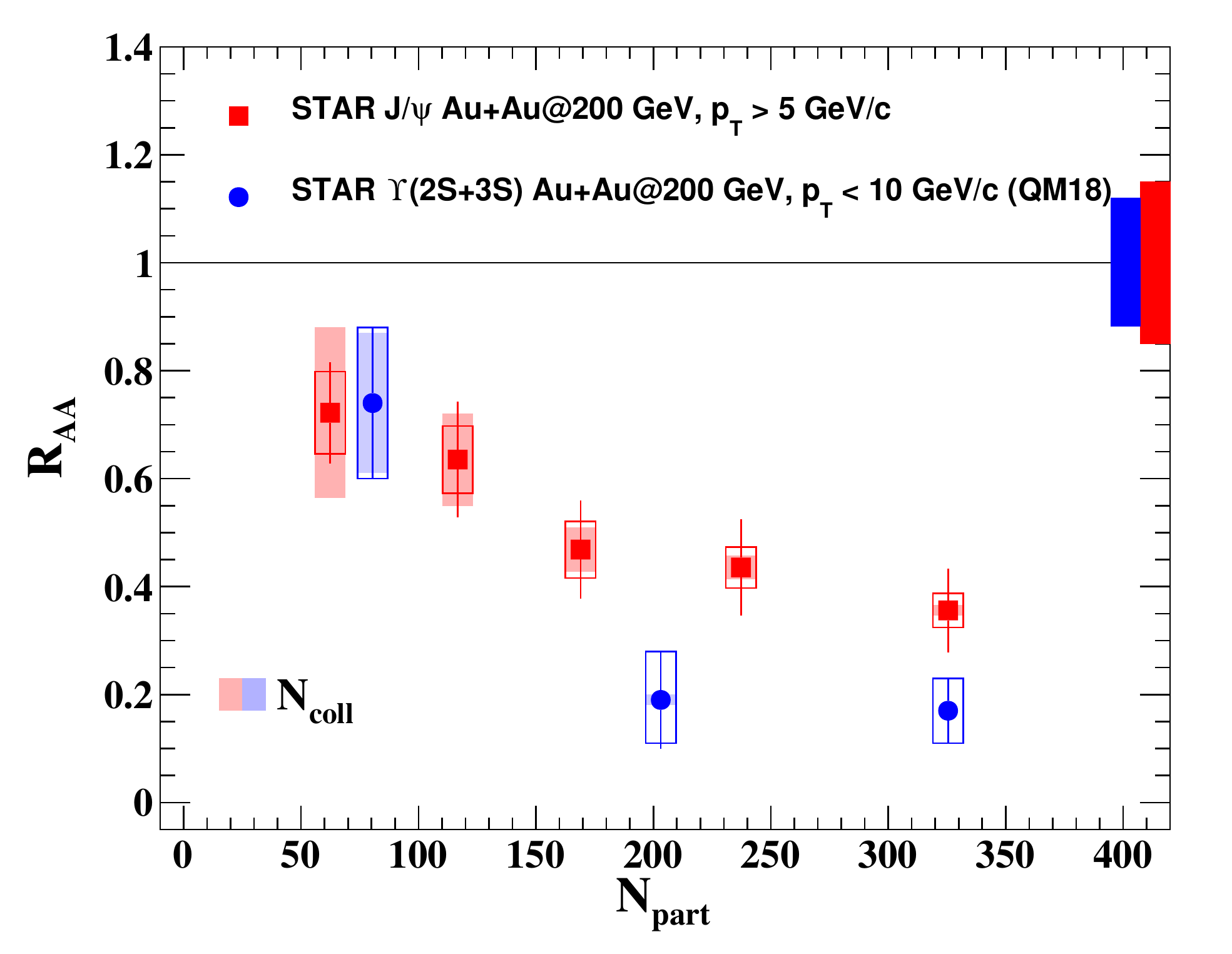}
  \includegraphics
  [width=0.32\hsize]
  {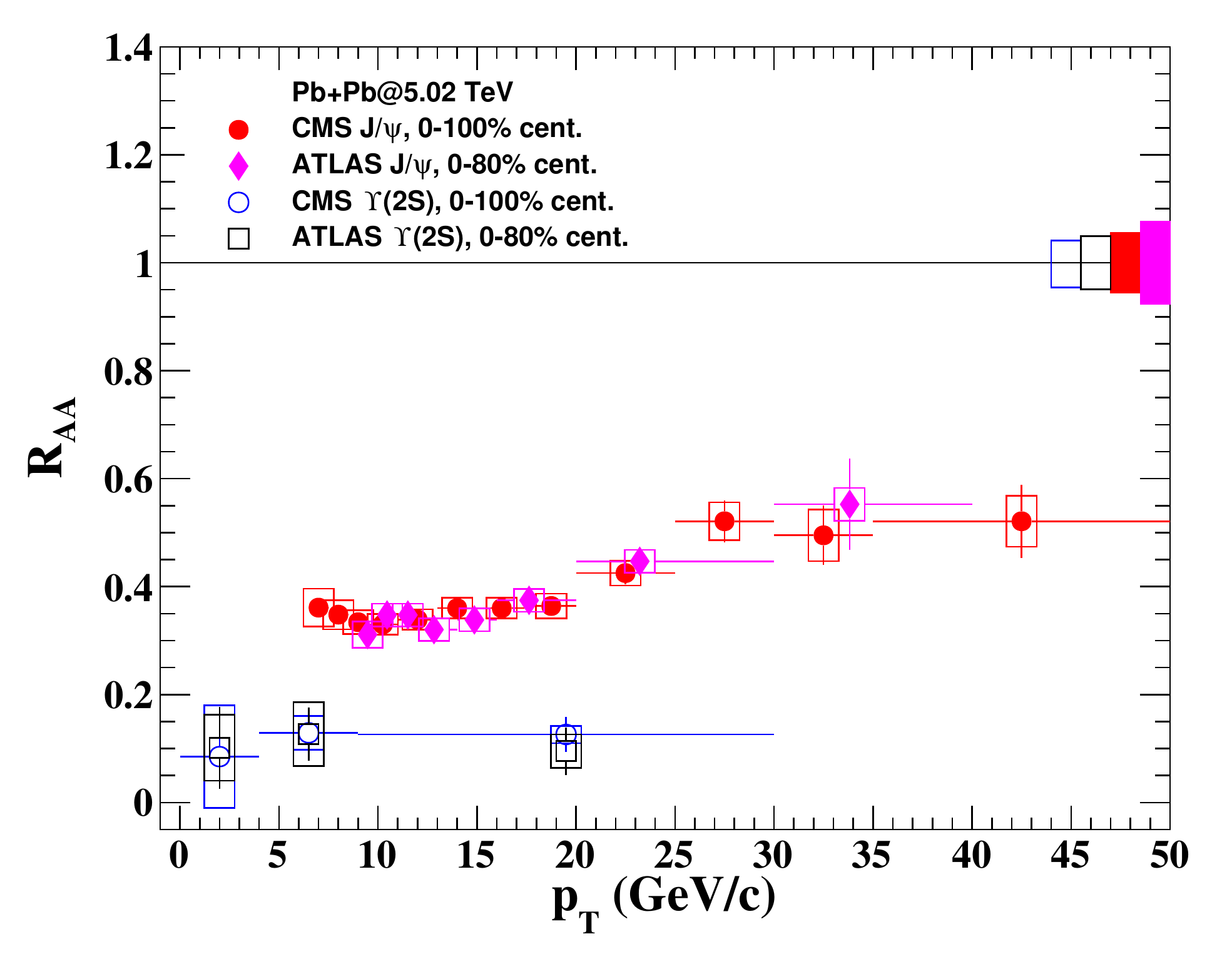}
  \includegraphics
  [width=0.32\hsize]
  {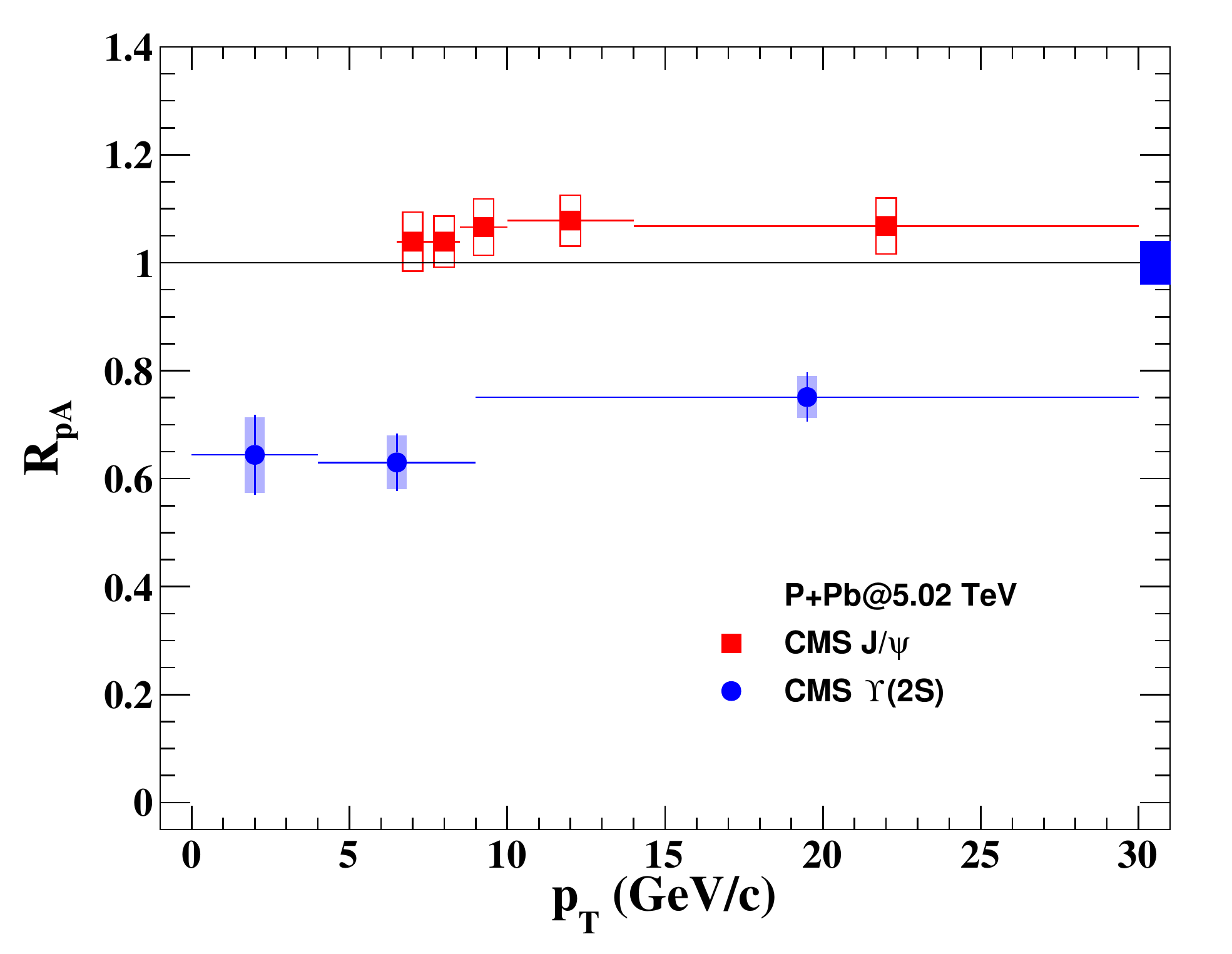}}
\caption{Comparison the nuclear modification factor of high-\pT \Jpsi and $\Upsilon(2S)$ (or $\Upsilon(2S+3S)$) in heavy-ion collisions and $p$+Pb collisions. }
\label{fig:Jpsi_Y2S}
\end{figure}

Since \Jpsi and $\Upsilon(2S)$ have similar radius, the suppression of high-\pT \Jpsi and $\Upsilon(2S)$ should be similar. However, the suppression of high-\pT \Jpsi is found to be much less than $\Upsilon(2S)$ at both RHIC and LHC as shown in the left two panels of Fig.~\ref{fig:Jpsi_Y2S}. This could not be explained by feed-down effect and possible (re)combination effect. As shown in the right panel of Fig.~\ref{fig:Jpsi_Y2S}, the nuclear modification factor of $\Upsilon(2S)$ is much smaller than that of \Jpsi at high-\pT in $p$+Pb collisions. The different CNM effects could be the reason for the different suppression measured in Pb+Pb collisions. Then the question is why the CNM effects of these two quarkonium states (with similar binding energy) are so different? Is it because of different nPDF effect? Nevertheless, it is important to consider CNM effects when interpreting $\Upsilon(2S)$ suppression measured in heavy-ion collisions, at least at LHC. 

\subsection{X(3872)}
X(3872) is the first exotic hadron discovered by the Belle Collaboration in 2003 in the $J/\psi \pi^+ \pi^-$ mass spectrum from $B$ decay~\cite{X3872_Belle2003}. The quantum numbers are measured to be incompatible with expected charmonium states~\cite{X3872_LHCb2013}. Its mass is larger than the sum of charm quark and anti-charm quark and is consistent with the sum of $D^0$ and $\overline{D^{*0}}$ masses. The internal structure of X(3872) is under debate, the possible candidates are tetraquark state or $D^0 \overline{D^{*0}}$ molecule state or mixed molecule-charmonium state. The tetraquark state and molecule state have very different binding energy. The radius of tetraquark is similar as that of charmonium state which is less than 1 $fm$. While for molecule state, the binding energy is very small, resulting in radius as large as several $fm$. Similar as quarkonium, the yield of X(3872) in dense QCD environment may shed light on the binding energy (internal structure). 

\begin{figure}[!htb] \centering{
\includegraphics
  [width=0.4\hsize, height=0.32\hsize]
  {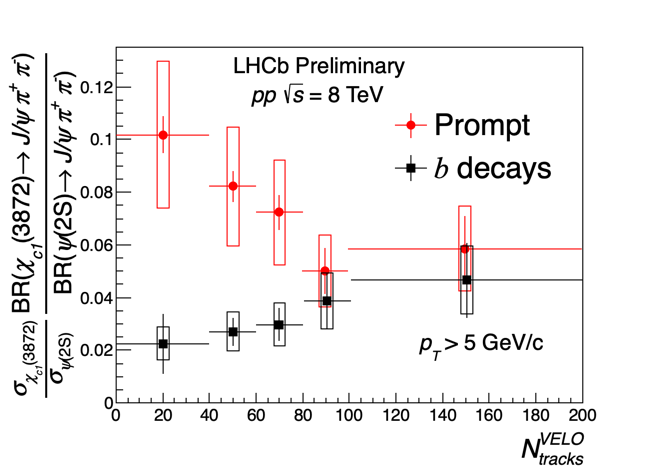}
  \includegraphics
  [width=0.4\hsize, height=0.32\hsize]
  {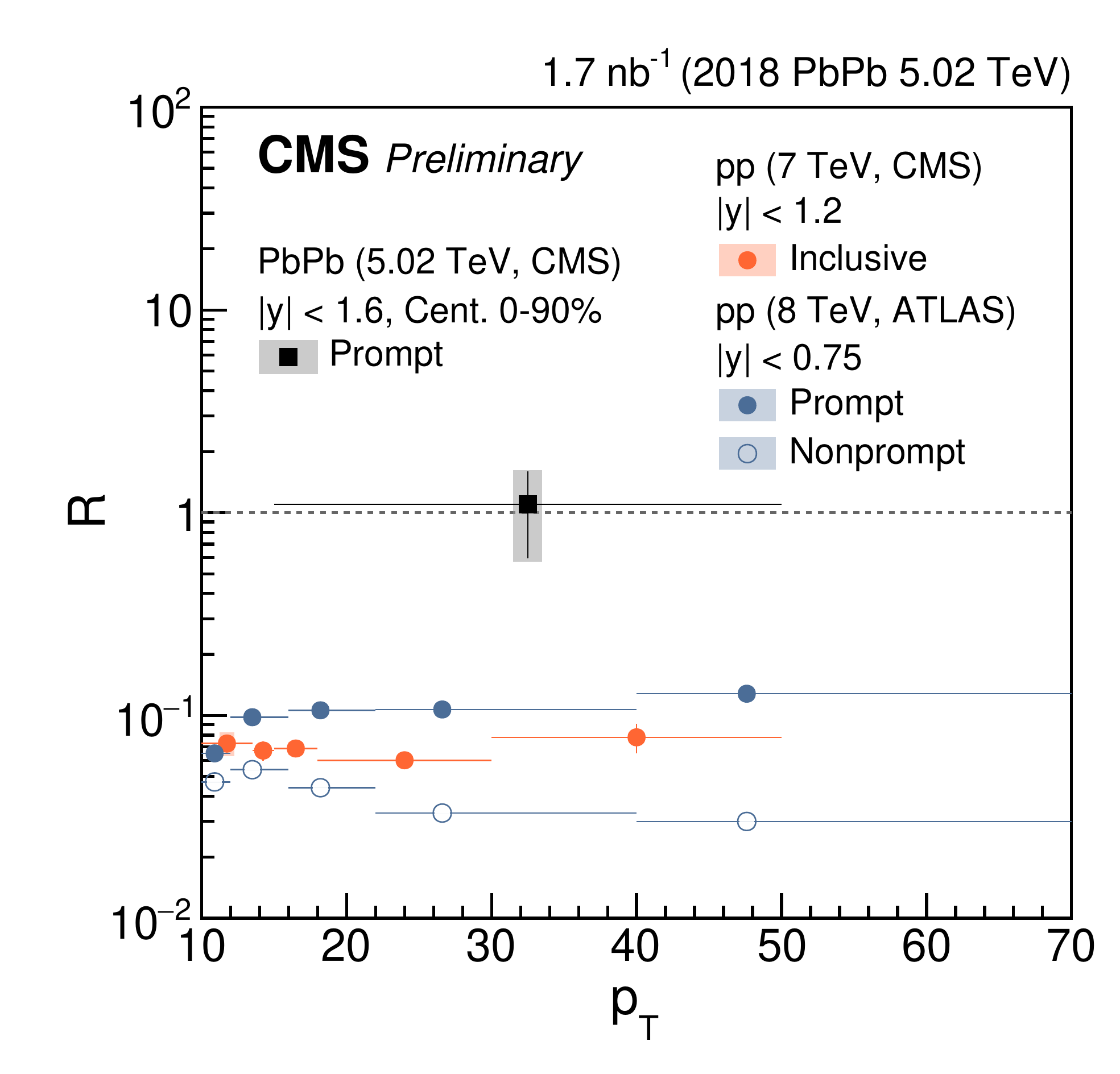}}
\caption{Left:  The cross section ratio of X(3872) over $\psi(2S)$ as a function of event activity in \pp collisions at \s = 8 TeV~\cite{X3872_LHCb_pPb}. Right:  The ratio as a function of \pT in Pb+Pb collisions at \sNN = 5.02 TeV~\cite{X3872_CMS_PbPb} compared to that in \pp collisions. }
\label{fig:X3872}
\end{figure}

The left panel of Fig.~\ref{fig:X3872} shows the ratio of high-\pT ($p_T > 5$ GeV/$c$) X(3872) (renamed to $\chi_{c1}(3872)$ by PDG) over $\psi(2S)$ as a function of event activity in \pp collisions at \s = 8 TeV for prompt and non-prompt component~\cite{X3872_LHCb_pPb}. It is found that the ratio has no significant change with event activity for non-prompt production. While for prompt production, there is increasing suppression of X(3872) relative to $\psi(2S)$ with increasing event activity. This is consistent with the interpretation of X(3872) as a large, weekly bound state. The CMS Collaboration measured the ratio of $1.1 \pm 0.51 (\textrm{stat.})  \pm 0.53 (\textrm{syst.})$ for high-\pT prompt X(3872) and $\psi(2S)$ in Pb+Pb collisions at \sNN = 5.02 TeV as shown in the right panel of Fig.~\ref{fig:X3872}~\cite{X3872_CMS_PbPb}. The central value is about a factor of 10 larger than that in \pp collisions, but the data is also consistent with \pp results within 1.5$\sigma$. Better precision is needed to extract the internal structure of the X(3872) particle.

\section{Conclusions}
In this paper, recent quarkonium(-like) measurements in $p$+$p$, $p$ or light nucleus + A, and A+A collisions at both RHIC and LHC are presented. In $p$ or light nucleus + A collisions, nPDF is important to explain the results, but additional effects are needed. At RHIC energy, nuclear absorption at backward rapidity is needed to bring down the nuclear modification factor for $J/\psi$. At LHC, nuclear absorption should be negligible due to small crossing time. However, final-state effects (such as breakup by comovers, possible hot medium effects) are needed especially at the Pb-going direction. In heavy-ion collisions, the quarkonium production at low and intermediate \pT fits the interplay of QGP melting, (re)combination and CNM effects. The (re)combination plays significant role in charmonium production at low and intermediate \pT at LHC. For $\Upsilon(2S)$ production in Pb+Pb collisions at LHC, CNM effects may play important role. At high-\pT region, \Jpsi may have sizable contribution from jet fragmentation. 

Some new observables such as \Jpsi polarization in heavy-ion collisions, $\Upsilon$ $v_2$ in heavy-ion collisions and X(3872) production yield in high multiplicity \pp collisions and Pb+Pb collisions are also presented. But better precision is needed to draw firm physics conclusion. 

\section*{Acknowledgement}
The author is supported in part by the National Key R\&D Program of China under Grand No. 2018YFE0104900, the National Natural Science Foundation of China under Grand Nos. 11675168 and 11720101001 and the Anhui Provincial Natural Science Foundation under Grand No. 1908085J02.


\bibliographystyle{elsarticle-num}
\bibliography{quarkonium_review.bib}







\end{document}